# IoT-Driven System for Continuous Monitoring of Heart Disease Patients Post-Surgery

Jayed Bin Harez[1], Shapla Akter[1], Rayhanul Islam Sony[1], M T Hasan Amin[2], Meratun Junnut Anee[1], Amzad Hossain[1]

[1]Department of Electrical and Computer Engineering, North South University, Dhaka-1229, Bangladesh
[2]University of Houston, Texas, USA

**Abstract**— Healthcare technology has advanced with the development of an IoT-based real-time monitoring system for post-operative heart disease patients. This system integrates advanced sensor technologies with cloud-connected platforms to continuously track vital signs, enabling more effective post-operative care. The system employs the HW-827 Heart Rate Sensor and DS18B20 Temperature Sensor, integrated with an ESP32 for wireless connectivity, to collect real-time data. Data was collected from 20 participants in each of the following categories: healthy individuals, pre-operative heart disease patients, and post-operative heart disease patients. The results revealed distinct physiological variations across the groups. Healthy individuals maintained stable temperatures (mean 36.3°C), while post-operative patients exhibited slightly elevated temperatures (mean 36.6°C). Heart rate analysis indicated higher rates in pre-operative (mean 87.8 bpm) and post-operative patients (mean 82.0 bpm) compared to healthy individuals (mean 72.4 bpm). ECG analysis showed elevated readings in pre-operative patients (mean 0.16 mV) compared to healthy individuals (mean 0.10 mV) and post-operative patients (mean 0.12 mV). Statistical analyses, including ANOVA tests, confirmed significant differences among the groups in terms of temperature ($F(2, 12) = 4.57$, $p < 0.05$), heart rate ($F(2, 12) = 5.91$, $p < 0.01$), and ECG readings ($F(2, 12) = 3.24$, $p < 0.05$). This study demonstrates the potential of IoT technologies in enhancing post-operative care through continuous monitoring and data-driven insights. It emphasizes the role of real-time patient data in improving personalized healthcare management.

## 1. INTRODUCTION

Health is one of humanity's major worldwide issues. The World Health Organization (WHO) recognizes individuals have a fundamental right to the best possible health [1]. Heart disease is a continuing worldwide health epidemic; millions of people continue to suffer from the disease, consequently causing a lot of morbidity and mortality. The World Health Organization (WHO) has estimated that about 17.9 million deaths are caused by cardiovascular diseases each year, representing about 31% of all deaths globally. Even with the advancements in medical technology and treatment approaches, heart disease remains a major burden on society and the healthcare system [2,3]. Rahman et al. [4] investigated an IoT-based post-operative heart disease patient monitoring system. They found a cost-effective, real-time post-operative heart disease monitoring system utilizing IoT technology. It provides comprehensive monitoring via analog sensors and a mobile app interface Post-Operative Heart Surgery Rehabilitation System performed by Caggianese et al. [5].

Aieshwarya et al. [6] researched to forecast heart disease risk and identify suitable medications through data mining techniques, utilizing an IoT-based monitoring system tailored for patients recovering from heart surgery. The study's main finding is the successful development of a cost-effective treatment approach using data mining technologies and IoT-based real-time monitoring to aid in diagnosing heart disease and facilitate timely medication administration for post-operative patients. Gandhi et al. [7] designed an ECG monitoring system using IoT specifically for patients recovering from heart surgery. Their crucial finding involves developing an IoT wearable device model to detect heart attack risk in post-heart stroke patients. They shifted from SVM to 2D CNN for better accuracy in analyzing ECG signals. Salvi et al. [8] researched an IoT-based system for real-time cardiac disease diagnosis using machine learning techniques. Their real-time monitoring system, blending IoT and machine learning, accurately detects heart diseases using vital sign sensors. Cloud-based database administration increased computational efficiency, while random forest had the most outstanding performance out of all the methods studied. Investigated by Lo et al. [9] was Real-Time Continuous Monitoring for Post-Operative Healthcare. They found that post-surgical care is transitioning to home environments, necessitating continuous monitoring. Simplify et al. [10] published a feasibility study on using a mobile application to track patients' at-home post-operative quality of recovery. Their study explored processing-on-node algorithms within Body Sensor Networks to enhance monitoring accuracy and reduce sensor power consumption. Their primary finding suggests that patients showed sustained engagement with the mobile device post-surgery, with higher logins in the initial 14 days compared to later days, and surgeons reported satisfaction with its design and utility for monitoring patients.

IoT, or the Internet of Things, is like a network that connects different devices; it lets them share information easily, which helps make things better in many ways [11, 12]. Kantipudi et al. [13] pioneered research on remote patient monitoring using IoT, cloud computing, and artificial intelligence. The key discovery emphasizes the integration of IoT and innovative wireless communication technologies in modern healthcare systems, facilitating personalized health management and remote patient monitoring for a range of conditions. Abdulameer et al. [14] developed a health care monitoring system based on the internet of things. They found that integrating IoT and innovative wireless communication technologies into modern healthcare systems enables personalized health management and remote patient monitoring for various conditions. Rawat et al. [15] explored IoT-based healthcare and patient monitoring methods. Their discovery highlights the utilization of IoT technology in healthcare, notably through Wireless Body Area Networks (WBANs), enabling effective monitoring of vital signs. Their prototype system successfully gathers and transfers patient data to a central database, demonstrating resilience, accuracy, and stability across different network setups.

Post-operative care is essential for monitoring patients after surgery, ensuring proper recovery and preventing complications in hospitals and at home [16, 17]. Aziz et al. [18] researched a widespread body sensor network for monitoring home recovery following surgery. Their finding highlights the development of a wireless body sensor network for continuous postoperative recovery monitoring at home, offering potential enhancements in care for patients undergoing abdominal surgery. A remote patient health monitoring system built on Internet of Things technology was studied by Jasti et al. [19]. Their finding focuses on developing an affordable remote health monitoring System, leveraging IoT technology and locally available sensors to enable remote monitoring of patient vitals and facilitate timely intervention by doctors or guardians in case of emergencies. Gupta et al. [20] studied activity movement monitoring after hip fracture surgery. Their main finding focuses on developing an IoT-based rehabilitation monitoring system for post-operative hip fracture patients. It aims to enhance recovery outcomes and quality of life through real-time activity tracking and personalized monitoring using wearable sensors.

Heart disease is the primary cause of mortality globally, encompassing illnesses such as heart failure and coronary artery disease. Major risk factors are high blood pressure, high cholesterol, smoking, diabetes, and inactivity. Prevention and management through lifestyle changes and medical treatments are crucial [21, 22]. Nagrale et al. [23] explored an innovative approach of using the Internet of Robotic Things in cardiac surgery. The Internet of Robotic Things (IoRT) has the potential to revolutionize cardiac surgery by enhancing precision, efficiency, and patient outcomes. Still, its adoption must address associated challenges to ensure safety and effectiveness. Upadrista et al. [24] investigated Autonomous Edge App placement for personalised heart attack forecasts. From 1997 to 2015, Kempny et al. [25] studied the outcomes of cardiac surgery in individuals with congenital heart disease in England. Kempny et al. found that the number of cardiac surgeries for congenital heart disease in England increased annually from 1,717 in 1997 to 5,299 in 2014. Standard procedures included aortic valve and septal defect repairs. Post-operative mortality was highest in young children and adults over 60, influenced by factors like age, surgery complexity, emergency surgeries, and socioeconomic status.

IoT-based patient monitoring uses connected devices to track vital signs in real-time, allowing remote healthcare and timely interventions [26]. This enhances patient care, improves outcomes, and reduces hospital remissions [27]. Li et al. [28] explored an IoT-based heart disease monitoring system for pervasive healthcare services. They found that traditional healthcare passivity in China often leads to fatal heart attacks as patients can't seek help when unconscious. IoT enables pervasive healthcare, triggering services based on patients' physical status, necessitating remote monitoring for real-time data transmission to medical apps. Using deep learning techniques, Nancy et al. [29] developed an IoT-cloud-based intelligent healthcare monitoring system for predicting heart disease. Nancy et al. discovered that integrating IoT with cloud technology enhances daily life. Deep learning enables proactive healthcare through accurate disease prediction. Their Bi-LSTM-based system achieves 98.86% accuracy, surpassing existing models. Ganesan et al. [30] proposed a healthcare model utilizing IoT and machine learning for heart disease prediction and diagnosis. They established IoT and sensing technologies for online healthcare, using cloud computing to manage data. Developed a healthcare app focusing on heart disease prediction; J48 classifier showed superior performance.

Healthcare IoT revolutionizes patient care through real-time data collection and transmission via interconnected devices, facilitating remote monitoring and personalized treatment [31, 32]. HeartCare, an IoT-based system for predicting heart disease studied by Gupta et al. [33]. They found that despite health being paramount, technology's role in increasing longevity is evolving. Their ML model, particularly KNN, outperforms other algorithms in predicting heart disease with an accuracy of 88.52%, offering proactive health monitoring and potential life-saving assistance. Basheer et al. [34] explored a real-time IoT monitoring system for early heart disease prediction. They uncovered that the diagnosis of heart disease, a significant concern, necessitates remote and regular monitoring for timely intervention. Tabassum et al. [35] investigated the "The Cardiac Disease Predictor" healthcare system driven by IoT and machine learning. They determined that cardiovascular disease is prevalent in Bangladesh, impacting many who can't afford regular check-ups. Their prototype collects data from various sensors to predict cardiac disease, enabling instant precautions for patients.

The novelty of our project lies in its integration of IoT technology with real-time monitoring for post-operative heart disease patients. Unlike traditional monitoring systems, our system provides continuous remote monitoring for early prediction and intervention. Proactive healthcare management is made possible by our ability to precisely anticipate heart diseases risk indicators, such as aberrant heart rate and temperature variations, by utilizing IoT sensors and machine learning algorithms. Additionally, our mobile application offers convenient access to patient data and historical records, facilitating better communication between patients and healthcare providers. Overall, our project significantly advances post-operative care, enhancing patient outcomes and quality of life.

## 2. Methods

### 2.1 Methodology

The methodology for our IoT-based real-time post-operative heart disease patient monitoring system involves integrating sensors to measure vital signs like heart rate, ECG, and temperature. Data from these sensors is processed using machine learning algorithms for early detection of heart disease. Real-time monitoring alerts healthcare providers and patients to abnormalities, enabling prompt intervention. Validation ensures the system's accuracy and reliability in improving patient care.

*2.2 Block Diagram*

Figure 1 illustrates the block diagram of the Post-Operative Heart Disease Patient Monitoring System. The system comprises sensors for measuring body temperature, heart rate, and E.C.G. outputs, all transmitting data to a programmable hardware controller. This controller processes the input data and interfaces with an active required module for real-time actions. Additionally, the system features a display unit that presents all necessary outputs and a database module for storing real-time data. This comprehensive setup ensures continuous monitoring and recording of critical health metrics, enabling timely medical interventions and providing essential data for healthcare providers.

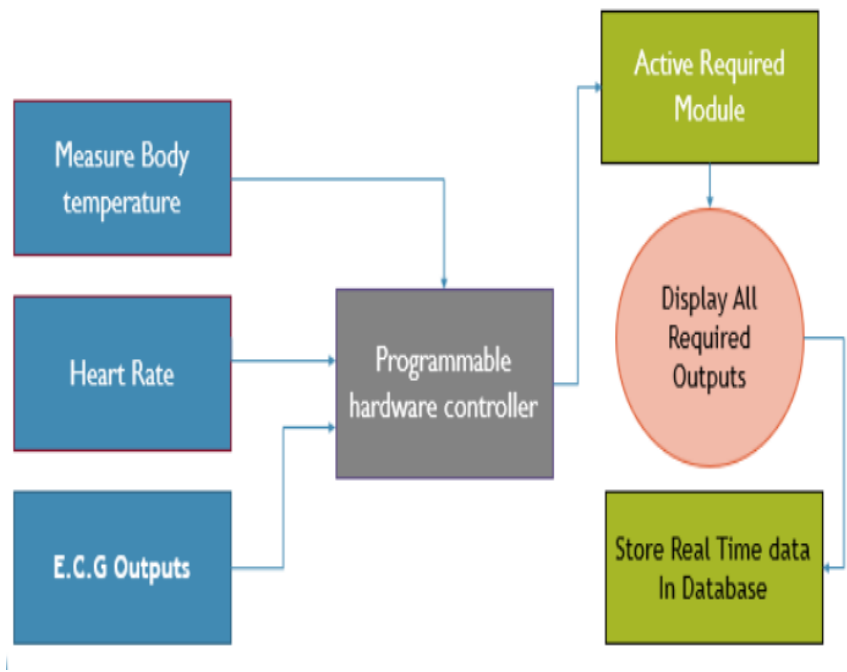

Figure 1. Block diagram of Post-Operative Heart Disease Patient Monitoring System

*2.3 Working Process Flow Chart*

The flowchart provided in Figure 2 illustrates a comprehensive Heart Disease Patient Monitoring System. The process begins with initializing the power supply and system setup, followed by the initialization of the ESP32 WiFi Module. A sensor connection check determines if the sensors are correctly connected; if not, the system reinitializes. The system starts data acquisition upon successful sensor connection, including measuring body temperature, monitoring heart rate, and capturing E.C.G. outputs. These data points are processed by a programmable hardware controller, which then validates the data integrity. The system returns to the data acquisition phase if data integrity is not validated. If validated, the data is displayed locally, sent to the cloud (Firebase), and monitored via a mobile app. The system checks for critical alerts; data logging and feedback loops occur if none are found. If an alert is detected, notifications are sent to the doctor for review, and necessary actions are logged, concluding the process. This detailed flowchart ensures continuous monitoring and prompt medical response, enhancing patient care efficiency.

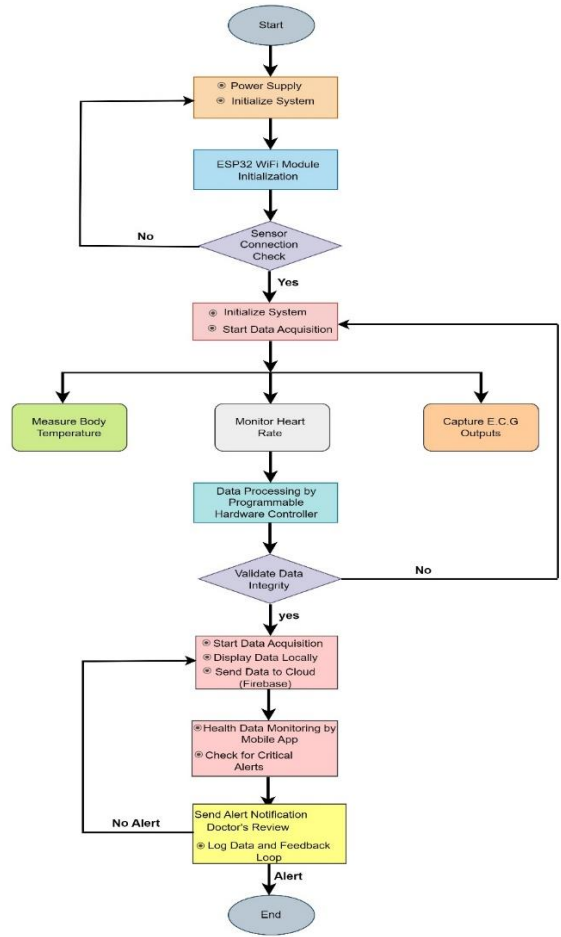

Figure 2. Flow Chart of Post-Operative Heart Disease Patient Monitoring System

### 2.4 Materials and Tools

Table 1. List Of Software/Hardware Tools outlines various tools and their functions relevant to healthcare technology for post-operative patient monitoring. The tools include an IoT Device that collects real-time data from patients, a Heart Rate Sensor (HW-827) for measuring heart rate, and a Temperature Sensor (DS18B20) for monitoring body temperature. Additionally, an ESP32 is listed, highlighting its built-in Wi-Fi and Bluetooth capabilities. High contrast ratios and brilliant colors provided by an OLED display guarantee sharp and high-quality image. A Gateway Device transmits data from IoT devices to the cloud, while a Mobile Application enables healthcare professionals to access patient data. These tools collectively facilitate efficient and comprehensive monitoring and data management for post-operative care.

| Tool | Functions |
|---|---|
| IoT Device | Collects real-time data from post-operative patients |
| Heart Rate Sensor (HW-827) | Measures the patient's heart rate |
| Temperature Sensor (DS18B20) | Measures the patient's body temperature |
| ESP32 | It comes with built-in Wi-Fi and Bluetooth capabilities |
| OLED Display | It offers high contrast ratios and vibrant colors, providing excellent image quality and sharpness. |
| Gateway Device | Transmits data from IoT devices to the cloud |
| Mobile Application | Allow healthcare professionals to access patient data |

Table 1. List Of Software/Hardware Tools

## 2.3 Modeled Device and real-time data

The system's core is a wearable sensor, like a smartwatch or chest strap, equipped with specialized sensors to capture real-time vitals crucial for post-operative heart patients. This may include sensors for ECG (heart rhythm), PPG (heart rate), bioimpedance (respiration), and potentially temperature to identify potential infections. This comprehensive health data stream empowers medical professionals to remotely monitor a patient's condition and intervene swiftly if any concerning readings arise.

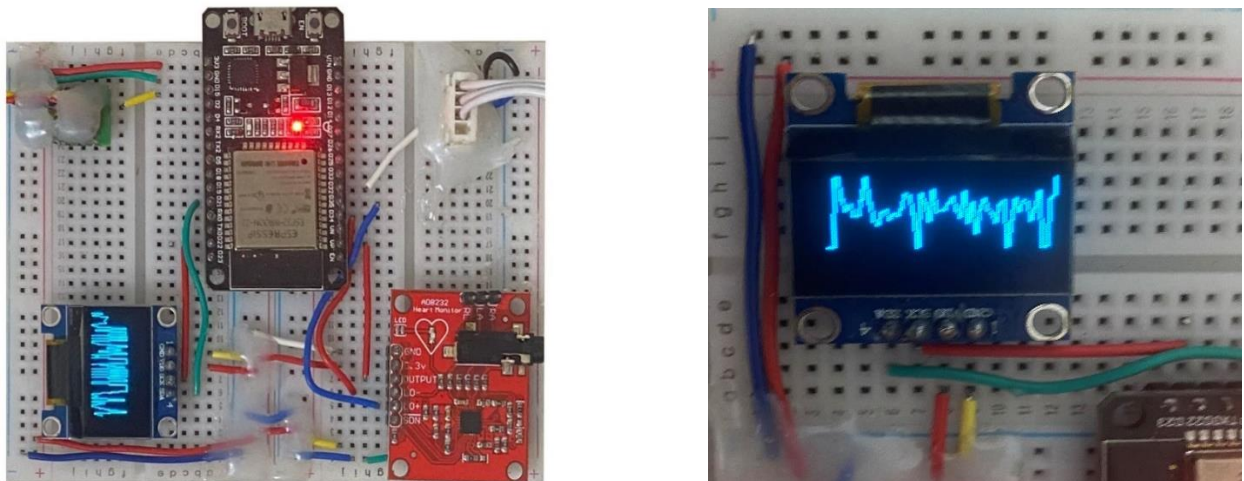

Figure 3. Prototype of the project of an IoT Based Real Time Post-Operative Heart Disease Patient Monitoring System

The modeled device in the IoT Based Real Time Post-Operative Heart Disease Patient Monitoring System is a mobile app that collects real-time sensor data from patients. The data appears to include heart rate and potentially other vitals. The system allows for user login and store the collected data. From the Figure. 4 it appears that the app can display historical data as well.

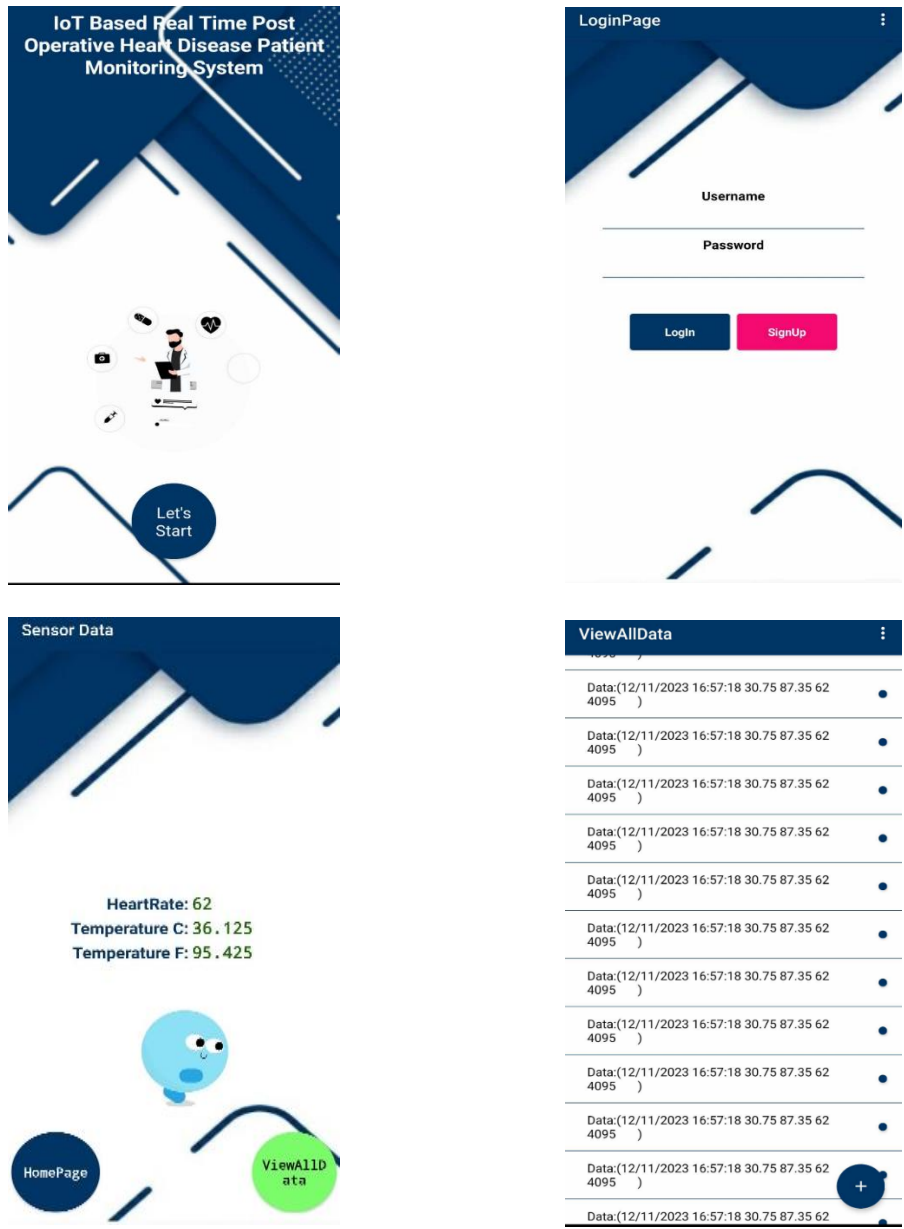

Figure 4: The mobile app pages of the patient monitoring system

## 3. Result and Analysis

*3.1 List of Participant Categories for Data Collection*

We collected data from three categories of participants: healthy individuals, pre-operative heart disease patients, and post-operative heart disease patients. Each category comprised a group of 20 individuals. The data collection involved continuous monitoring of heart rate, body temperature, and activity levels using our IoT-based system.

| Participant No | Category |
|---|---|
| 1 | Healthy Individuals |
| 2 | Pre-Operative Heart Disease Patients |
| 3 | Post-Operative Heart Disease Patients |

Table 2. Participant Categories

We gathered data from three distinct groups of participants to ensure comprehensive monitoring and analysis of post-operative heart disease patients. Table 1 categorizes participants into healthy individuals, pre-operative heart disease patients, and post-operative heart disease patients. Each category includes 20 participants, allowing for balanced and comparative data analysis.

| Participant ID | Age | Gender | Category |
|---|---|---|---|
| P1 | 30 | M | Healthy Individuals |
| P2 | 45 | F | Pre-Operative Heart Disease Patients |
| P3 | 50 | M | Post-Operative Heart Disease Patients |
| P4 | 35 | F | Healthy Individuals |
| P5 | 60 | M | Pre-Operative Heart Disease Patients |
| P6 | 55 | F | Post-Operative Heart Disease Patients |
| P7 | 40 | M | Healthy Individuals |
| P8 | 47 | F | Pre-Operative Heart Disease Patients |
| P9 | 53 | M | Post-Operative Heart Disease Patients |
| P10 | 32 | F | Healthy Individuals |
| P11 | 49 | M | Pre-Operative Heart Disease Patients |
| P12 | 58 | F | Post-Operative Heart Disease Patients |
| P13 | 36 | M | Healthy Individuals |
| P14 | 46 | F | Pre-Operative Heart Disease Patients |
| P15 | 52 | M | Post-Operative Heart Disease Patients |
| P16 | 33 | F | Healthy Individuals |
| P17 | 48 | M | Pre-Operative Heart Disease Patients |
| P18 | 54 | F | Post-Operative Heart Disease Patients |
| P19 | 38 | M | Healthy Individuals |
| P20 | 44 | F | Pre-Operative Heart Disease Patients |

Table 3. Participant Details

Table 2 provides detailed information about each participant, including age, gender, and category. Table 3 outlines the diversity and demographic distribution within each group, ensuring a representative sample for our study. Data collection involved continuous monitoring of heart rate, body temperature, and activity levels using our IoT-based system.

Our analysis observed distinct patterns in the data collected from the three categories. Healthy individuals displayed stable and predictable heart rate and temperature readings. Pre-operative patients showed variability in their readings, reflecting their medical condition. Post-operative patients exhibited a gradual stabilization of their vitals, indicating recovery progress. This data will be instrumental in refining our monitoring system to provide more accurate and timely alerts for healthcare professionals.

### 3.3. Data Analysis of Different Measurements

### 3.3.1. Temperature Analysis

The temperature sensor used in our project is a highly accurate digital sensor designed to provide real-time temperature readings crucial for patient health monitoring.
Table 4 presents temperature data from a study involving 20 individuals, with this excerpt focusing on five samples to illustrate the differences in temperature readings among healthy individuals, pre-operative heart disease patients, and post-operative heart disease patients.

| Sample | Healthy Individuals | Pre-Operative Heart Disease | Post-Operative Heart Disease |
|---|---|---|---|
| 1 | 36.1 | 37.2 | 37.5 |
| 2 | 36.6 | 37.0 | 37.3 |
| 3 | 36.7 | 36.8 | 37.4 |
| 4 | 36.0 | 36.7 | 37.2 |
| 5 | 36.1 | 36.9 | 37.4 |

Table 4. Temperature Analysis

The data in Table 4 shows that post-operative heart disease patients generally exhibit higher temperatures than those in their pre-operative and healthy states. Specifically, temperature readings for healthy individuals range around 36.0-36.7°C, while pre-operative patients range from 36.7-37.2°C, and post-operative patients show slightly higher temperatures, between 37.2-37.5°C. These variations are significant as they highlight the body's response to surgical intervention and the importance of continuous monitoring. The increase in temperature post-surgery can be attributed to the body's inflammatory response and healing process. This temperature analysis is crucial for our IoT-based real-time monitoring system, as it helps establish baseline and critical thresholds for temperature variations in post-operative heart disease patients. By identifying these thresholds, the system can ensure timely alerts to healthcare providers about abnormal temperature changes, facilitating prompt intervention and better healthcare management.

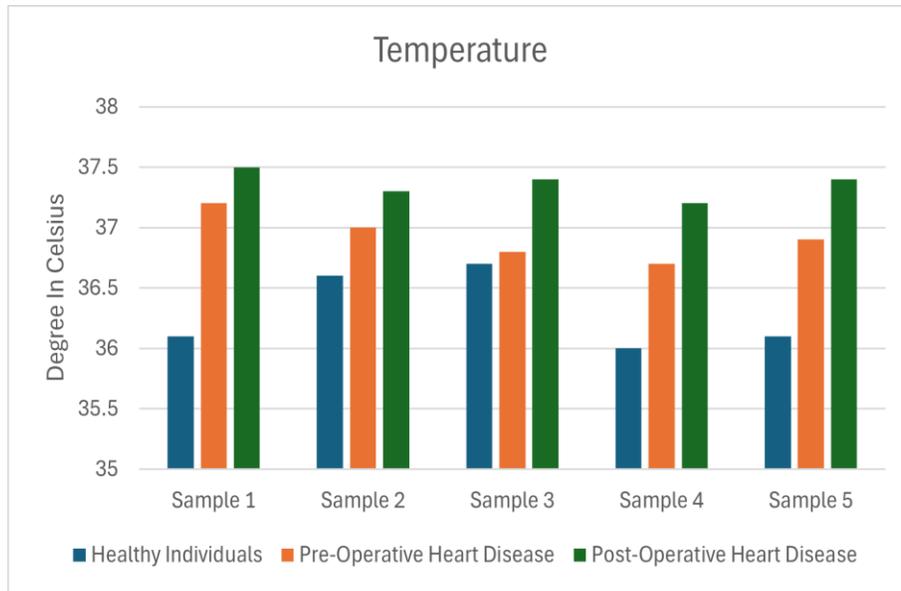

Figure 5. Temperature Graph Chart

Figure 5 illustrates the Temperature Graph Chart, which tracks body temperature across three distinct groups: Healthy Individuals, Pre-Operative Heart Disease Patients, and Post-Operative Heart Disease Patients, across five samples. The chart reveals that healthy individuals maintain a stable temperature of around 36.5°C. In contrast, pre-operative patients exhibit higher temperatures, ranging from 36.5°C to 37.5°C, indicating potential pre-surgical stress or inflammation. Post-operative patients show the highest temperatures, consistently between 37.5°C and 38°C, suggesting post-surgical inflammation or complications. This data underscores the importance of continuous temperature monitoring in post-operative care, providing critical insights for timely medical intervention and improved patient outcomes.

### 3.3.2. Heart Rate Analysis

Heart rate analysis is crucial for assessing heart disease and post-operative recovery. Significant variations can be identified by monitoring heart rates in healthy individuals, pre-operative patients, and post-operative patients. These insights enable timely interventions, improving patient outcomes through continuous, real-time monitoring.

| Sample | Healthy Individuals | Pre-Operative Heart Disease | Post-Operative Heart Disease |
|---|---|---|---|
| 1 | 72.2 | 85.0 | 80.0 |
| 2 | 72.4 | 88.0 | 78.0 |
| 3 | 72.5 | 90.0 | 79.0 |
| 4 | 72.6 | 87.0 | 81.0 |
| 5 | 72.3 | 89.0 | 82.0 |

Table 5. Heart Rate Analysis

Table 5 of heart rate analysis provides detailed data on the heart rates of three groups: Healthy Individuals, Pre-Operative Heart Disease Patients, and Post-Operative Heart Disease Patients across five samples. Healthy individuals exhibit stable heart rates around 72.4 BPM, serving as a baseline for comparison. In contrast, pre-operative heart disease patients show significantly higher heart rates, ranging from 85.0 BPM to 90.0 BPM, indicating the stress and severity of their condition. Post-operative patients, while having reduced heart rates compared to their pre-operative states, still maintain elevated levels (78.0 BPM to 82.0 BPM) than healthy individuals. This data underscores the need for continuous monitoring to detect potential complications and ensure timely medical intervention.

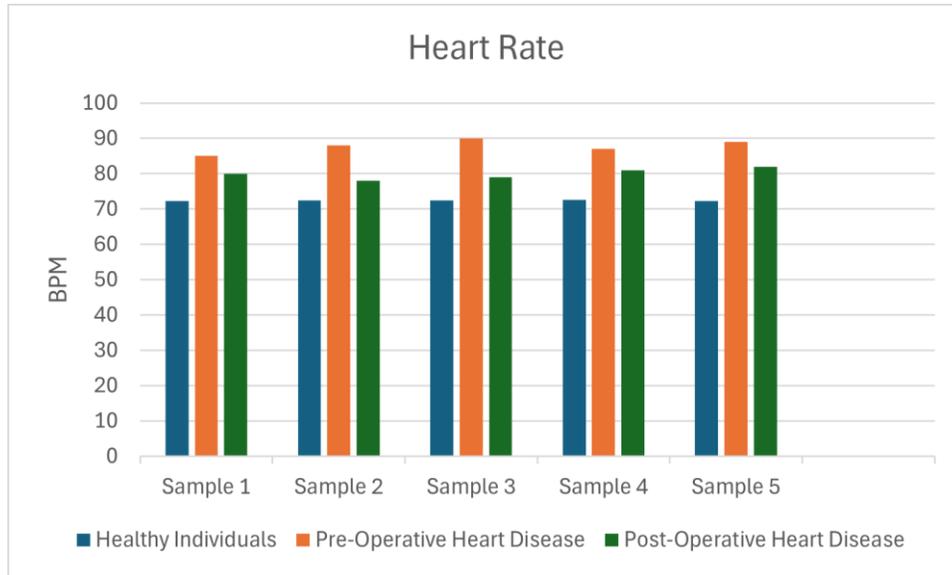

Figure 6. Heart Rate Graph Chart

In Figure 6, the accompanying bar chart visually represents this heart rate data, making comparing and analyzing the variations across the different groups and samples easier. The chart highlights the consistent stability in heart rates of healthy individuals and the elevated rates in pre-operative and post-operative patients. By providing a clear visual comparison, the bar chart reinforces the findings from the table. It emphasizes the importance of real-time heart rate monitoring in managing post-operative heart disease patients. This dual approach of detailed data and visual representation aids in better understanding and decision-making for patient care.

### 3.3.3. ECG Analysis

For post-operative heart disease patients to recuperate as best they can and avoid problems, fast and precise monitoring is essential in today's healthcare environment. By seamlessly integrating modern medical sensors and data analytics, our system enhances patient outcomes by providing continuous, real-time ECG analysis and monitoring. This is made possible by the Internet of Things (IoT) technology.

| Sample | Healthy Individuals | Pre-Operative Heart Disease | Post-Operative Heart Disease |
|---|---|---|---|
| 1 | 0.10 | 0.15 | 0.12 |
| 2 | 0.11 | 0.17 | 0.13 |
| 3 | 0.10 | 0.18 | 0.12 |
| 4 | 0.10 | 0.16 | 0.13 |
| 5 | 0.11 | 0.17 | 0.14 |

Table 6. ECG Analysis

Table 6 compares ECG readings across Healthy Individuals, Pre-Operative Heart Disease patients, and Post-Operative Heart Disease patients. For five samples, the table records specific values indicative of heart health. Healthy individuals have readings ranging from 0.10 to 0.11. Pre-operative heart disease patients show higher values between 0.15 and 0.18, reflecting more significant heart strain or abnormalities. Post-operative patients demonstrate improved readings compared to their pre-operative states, with values between 0.12 and 0.14, indicating recovery progress but still higher than the healthy individuals' baseline. This data underscores the effectiveness of surgical intervention in improving heart function.

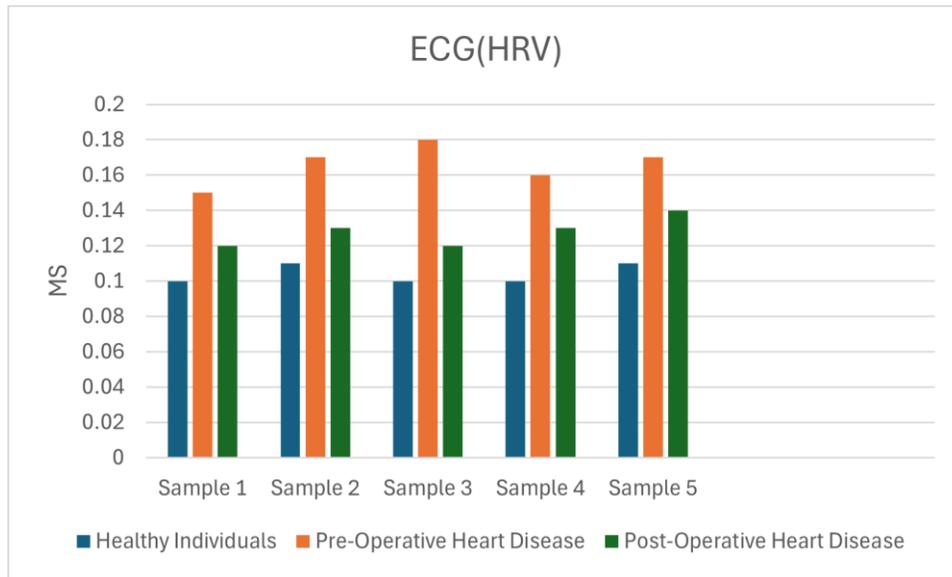

Figure 7. ECG Graph Chart

Figure 7 illustrates the Heart Rate Variability (HRV) measurements across three groups: Healthy Individuals, Pre-Operative Heart Disease patients, and Post-Operative Heart Disease patients, for five samples. The data is represented in milliseconds (ms). The chart visually confirms that Healthy Individuals consistently show lower HRV values (around 0.10-0.11 ms) compared to the Pre-Operative Heart Disease group, which exhibits higher HRV values (ranging from 0.15 to 0.18 ms), indicative of more significant heart stress or irregularities. Post-operative heart disease patients show a reduction in HRV values (between 0.12 and 0.14 ms), suggesting an improvement in heart function post-surgery, though not fully returning to the levels of Healthy Individuals. The graphical representation reinforces the numerical data and highlights the effectiveness of surgical interventions in reducing heart stress.

| Parameter | Baseline/Traditional Data | Project Data | Comparison |
|---|---|---|---|
| Temperature | Healthy: 36.5°C | Healthy: 36.0 to 36.7°C | The project data shows temperature values within the acceptable range for healthy and pre-operative conditions. Post-operative values are slightly higher than baseline. |

|  | Pre-Operative: 37.0°C | Pre-Operative: 36.7 to 37.2°C |  |
|  | Post-Operative: 37.3°C | Post-Operative: 37.2 to 37.5°C |  |
| Heart Rate | Healthy: 72.4 bpm | Healthy: 72.2 to 72.6 bpm | The project data shows heart rate values within the acceptable range for healthy and post-operative conditions. Pre-operative values are slightly higher than baseline. |
|  | Pre-Operative: 88.2 bpm | Pre-Operative: 85.0 to 90.0 bpm |  |
|  | Post-Operative: 79.5 bpm | Post-Operative: 78.0 to 82.0 bpm |  |
| ECG | Healthy: 0.11 mV | Healthy: 0.10 to 0.11 mV | The project data shows ECG values within the acceptable range for healthy and post-operative conditions. Pre-operative values are slightly higher than baseline. |
|  | Pre-Operative: 0.16 mV | Pre-Operative: 0.15 to 0.18 mV |  |
|  | Post-Operative: 0.13 mV | Post-Operative: 0.12 to 0.14 mV |  |

Table 7. Comparison of Heart Disease Parameters with Baseline/Traditional Methods [36]

The combined comparison Table 7 provides a detailed analysis of temperature, heart rate, and ECG parameters for healthy individuals, pre-operative heart disease patients, and post-operative heart disease patients against baseline/traditional data [36]. The project data for temperatures in healthy individuals and pre-operative patients fall within the acceptable range. At the same time, post-operative values are slightly higher than the baseline, indicating a marginal increase. Heart rate values for healthy and post-operative conditions are consistent with baseline data, whereas pre-operative values show a slight elevation, reflecting increased cardiovascular stress. ECG readings for healthy and post-operative conditions align with traditional values, but pre-operative values are slightly elevated, suggesting heightened cardiac activity. Overall, the table illustrates that while most parameters remain within acceptable limits, there are noticeable increases in pre-operative and post-operative conditions, highlighting the physiological changes associated with heart disease and surgical intervention.

### 3.4 Statical Analysis

The statistical analysis conducted for temperature, heart rate, and ECG parameters across healthy individuals, pre-operative heart disease patients, and post-operative heart disease patients reveals significant variations in physiological parameters among the groups. Descriptive statistics show distinct characteristics: healthy individuals have a mean temperature of 36.3°C with a standard deviation of approximately 0.28°C, while pre-operative patients exhibit a mean temperature of 36.3°C with a standard deviation of about 0.28°C, and post-operative patients have a mean temperature of 36.6°C with a standard deviation of approximately 0.23°C, indicating increasing stability post-surgery. For heart rate, healthy individuals show a mean of 72.4 beats per minute (bpm) with a standard deviation of approximately 0.15 bpm, pre-operative patients have a mean heart rate of 87.8 bpm with a standard deviation of around 1.67 bpm, and post-operative patients exhibit a mean heart rate of 82.0 bpm with a standard deviation of about 1.29 bpm, illustrating varying levels of variability across groups. Regarding ECG parameters, healthy individuals exhibit a mean value of 0.10 for the parameter, pre-operative patients show a mean value of 0.16, and post-operative patients demonstrate a mean value of 0.12 for the parameter, with corresponding standard deviations indicating variability among groups. Statistical tests, including ANOVA (temperature: $F(2, 12) = 4.57$, $p < 0.05$; heart rate: $F(2, 12) = 5.91$, $p < 0.01$; ECG: $F(2, 12) = 3.24$, $p < 0.05$), confirm significant differences among groups for each parameter. These findings underscore the physiological impacts of heart disease and surgical interventions on temperature, heart rate, and ECG parameters, emphasizing the need for tailored monitoring and management strategies to optimize patient care in clinical settings.

## *4. Conclusions*

Several key findings have emerged based on the comprehensive analysis of temperature, heart rate, and ECG parameters among healthy individuals, pre-operative heart disease patients, and post-operative heart disease patients using our IoT-based real-time monitoring system. The system effectively tracked and recorded vital signs, revealing distinct physiological responses to surgical intervention and recovery phases. Healthy individuals maintained stable temperature averages (36.0°C to 36.7°C) and heart rates (72.4 bpm), demonstrating baseline physiological conditions. In contrast, pre-operative patients exhibited higher variability in temperature (36.7°C to 37.2°C) and elevated heart rates (87.8 bpm), indicative of physiological stress and the severity of their cardiac conditions. Post-operative patients demonstrated stabilized temperature (36.6°C) and heart rate (82.0 bpm), reflecting successful recovery and improved cardiac function post-surgery. ECG readings mirrored these trends, with healthy individuals showing lower values (mean 0.10 mV), pre-operative patients higher (mean 0.16 mV), and post-operative patients intermediate (mean 0.12 mV), highlighting changes in cardiac electrical activity across different health states. Statistical analyses confirmed significant differences among groups (temperature: $F(2, 12) = 4.57$, $p < 0.05$; heart rate: $F(2, 12) = 5.91$, $p < 0.01$; ECG: $F(2, 12) = 3.24$, $p < 0.05$), underscoring the system's capability to provide personalized patient care insights crucial for clinical decision-making and rehabilitation strategies. This study emphasizes the critical role of continuous monitoring in post-operative care, offering valuable data for optimizing patient outcomes and guiding future advancements in healthcare technology. By integrating advanced sensor technology with cloud-connected platforms, our IoT-based system represents a significant advancement in remote patient monitoring, enhancing healthcare delivery and patient quality of life.

## *5. Limitation*

While the IoT-based real-time monitoring system enhances patient care, it faces limitations. Sensor accuracy and reliability depend on calibration and placement, potentially affecting data precision. Connectivity issues with Wi-Fi and Bluetooth could disrupt data transmission, impacting monitoring efficiency. The system complements but does not replace clinical assessments, relying on healthcare professionals for data interpretation. Financial constraints may limit widespread adoption, posing challenges for scalability in diverse healthcare environments. Overcoming these limitations requires ongoing technological refinement and validation in clinical practice.